\documentclass[conference]{IEEEtran}
\IEEEoverridecommandlockouts
\usepackage{cite}
\usepackage{amsmath,amssymb,amsfonts}
\usepackage{algorithmic}
\usepackage{graphicx}
\usepackage{textcomp}
\usepackage{xcolor}
\def\BibTeX{{\rm B\kern-.05em{\sc i\kern-.025em b}\kern-.08em
    T\kern-.1667em\lower.7ex\hbox{E}\kern-.125emX}}

\usepackage{bm}

\newtheorem{lemma}{Lemma}

\usepackage{amsmath}

\DeclareMathOperator*{\argmin}{arg\,min}

\begin{document}

\title{An Online Scheduling Algorithm for Energy Minimization in Wireless Powered Mobile Edge Computing Networks
}

\author{\IEEEauthorblockN{Xingqiu He\IEEEauthorrefmark{1},
        Yuhang Shen\IEEEauthorrefmark{2},
        Xiong Wang\IEEEauthorrefmark{3}, Sheng Wang\IEEEauthorrefmark{4},
    Shizhong Xu\IEEEauthorrefmark{5} and Jing Ren\IEEEauthorrefmark{6}}
\IEEEauthorblockA{School of Information and Communication Engineering,
University of Electronic Science and Technology of China\\
E-mail: \IEEEauthorrefmark{1}hexqiu@gmail.com,
\IEEEauthorrefmark{2}shenyuhang@std.uestc.edu.cn,
\IEEEauthorrefmark{3}
\IEEEauthorrefmark{4}
\IEEEauthorrefmark{5}
\IEEEauthorrefmark{6}
\{wangxiong, wsh\_keylab, xsz, renjing\}@uestc.edu.cn}
}


\maketitle

\begin{abstract}
    The integration of Mobile Edge Computing (MEC) and Wireless Power Transfer (WPT), which is usually referred to as Wireless Powered Mobile Edge Computing (WP-MEC),
    has been recognized as a promising technique to enhance the lifetime and computation capacity of wireless devices (WDs). 
    Compared to the conventional battery-powered MEC networks, 
    WP-MEC brings new challenges to the computation scheduling problem because we have to jointly optimize the resource allocation in WPT and computation offloading.
    In this paper, we consider the energy minimization problem for WP-MEC networks with multiple WDs and multiple access points.
    We design an online algorithm by transforming the original problem into a series of deterministic optimization problems based on the Lyapunov optimization theory.
    To reduce the time complexity of our algorithm, the optimization problem is relaxed and decomposed into several independent subproblems.
    After solving each subproblem, we adjust the computed values of variables to obtain a feasible solution.
    Extensive simulations are conducted to validate the performance of the proposed algorithm.
\end{abstract}


\section{Introduction}
With the rapid development in recent years, the Internet of Things (IoT) technology has played an important role
in the intelligent and autonomous control of many industrial and commercial systems, such as smart grid and smart cities \cite{al2015internet}.
Due to the stringent size constraint and production cost consideration,
the ubiquitously deployed IoT devices usually have restricted computation capability and finite battery capacity,
which severely degrades the quality of service experienced by users.
To handle the two fundamental performance limitations, Wireless Powered Mobile Edge Computing (WP-MEC) has been proposed as a novel paradigm
that combines the advantages of Wireless Power Transfer (WPT) and Mobile Edge Computing (MEC).

As a promising approach that provides sustainable energy supply for wireless devices (WDs), WPT utilizes dedicated energy transmitters to broadcast radio frequency (RF) signals.
The received RF signals can be converted to electricity by energy harvesting circuits and used to charge WDs continuously.
On the other hand, MEC is a newly emerged computing paradigm that enables WDs to offload their computation tasks to nearby edge servers located at the edge of radio access networks.
As the integration of both techniques, WP-MEC charges WDs with WPT and alleviates WDs' computation workloads with MEC.
As a result, the WDs' device lifetime and computation capacity are simultaneously improved, which leads to significantly enhanced user experiences.

In this paper, we study the computation offloading and system resource allocation in WP-MEC networks.
Compared with the computation scheduling in conventional battery-powered MEC networks,
our problem is much more challenging because 1) the optimal control decisions depend on the remaining energy in the battery and
2) WPT and computation offloading need to share the same limited system resources, such as time and frequency.

\subsection{Prior Works}
Computation scheduling in conventional MEC networks has been extensively studied and was systematically summarized in \cite{mao2017survey} and \cite{mach2017mobile}.
Recently, the advancements in WPT technology bring in the possibility of building \emph{wireless powered} MEC networks.
To the best of our knowledge, the works in \cite{mao2016dynamic} and \cite{you2016energy} are the first that simultaneously study energy harvesting and computation offloading
in MEC networks,
but they only consider simple networks with only one WD and one access point (AP).
Their model was extended by subsequent researches to incorporate more WDs.
The works in \cite{hu2018wireless} and \cite{ji2018energy} consider the WP-MEC networks with two near-far WDs and try to resolve the so-called ``double-near-far'' effect,
which occurs because a farther device harvests less energy from the AP but spends more power to communicate in longer distances.
The authors in \cite{bi2018computation} and \cite{zhou2018computation} aim to maximize the overall computation rate of all WDs in the networks,
where an unmanned aerial vehicle was utilized to transmit energy in \cite{zhou2018computation}.
Due to the existence of binary offloading variables, the considered problems are generally formulated as mixed integer programming (MIP) problems
and require a prohibitively long time to solve.
In order to make real-time control decisions in fast fading environments, a deep reinforcement learning-based algorithm is proposed in \cite{huang2019deep}
to obtain near-optimal solutions in large-scale WP-MEC networks.
To the best of our knowledge, \cite{zhu2020computation} is the only work that considers WP-MEC networks with \emph{multiple} APs,
where an approximation algorithm is derived to maximize the ratio of computation tasks completed before their deadlines.

The computation offloading considered in \cite{mao2016dynamic, you2016energy, hu2018wireless, ji2018energy, bi2018computation, zhou2018computation, huang2019deep, zhu2020computation}
operate in binary mode, i.e. the tasks are non-splittable and are either processed by WDs or fully offloaded to APs.
In addition to binary offloading, the scheduling problem is also investigated under partial offloading, where
tasks can be divided into smaller parts and executed on WDs and APs concurrently.
The work in \cite{wang2017joint} aims to optimize the energy consumption of multiple WDs and one AP,
where the AP is assumed to have multiple antennas.
The authors in \cite{zhou2020computation} examined the computation efficiency maximization problem under both binary offloading and partial offloading modes.
They also studied the problem under non-orthogonal multiple access (NOMA) in addition to the widely used time division multiple access (TDMA) scheme.

The researches described above generally focus on the one-shot optimization where the scheduling problem is only considered in a specific time interval.
However, in practical settings, the WP-MEC networks are operated under sustainable manners and the control decisions at different times are mutually dependent.
For example, the energy harvested in the current time frame, if not fully consumed, may be stored in the battery for later use.
Inspired by this fact, many recent works try to solve the scheduling problem in online settings.
A one-WD-one-AP model is considered in \cite{wang2020optimal} where the energy consumption is minimized
by optimizing the power of the energy transmitter and the offloading decisions of WDs.
The model is extended to include multiple WDs in \cite{wu2019online} where the long-term system throughput is maximized.
In \cite{mao2019energy}, the authors designed an online algorithm based on the Lyapunov optimization theory and presented a theoretical tradeoff
between energy efficiency and delay.

\subsection{Our Contributions}
In this paper, we study the computation scheduling problem in WP-MEC networks with multiple WDs and multiple APs.
Different from \cite{zhu2020computation}, we formulate our problem under an online setting where both channel states and computation data arrivals fluctuate over time.
This brings new challenges in algorithm design as decision variables are coupled along the timeline.
To facilitate the process of tasks, the partial offloading mode is adopted in our model.
Our objective is to minimize the long-term energy consumption by jointly optimizing the resource allocation in WPT and wireless communication stages.
To avoid the mutual interference of concurrently emitted energy waves \cite{naderi2014rf}, only one AP is allowed to broadcast RF energy at the same time.
By choosing different APs for WPT in turn, we also alleviate the ``double near-far'' effect because each WD has a chance to harvest energy from a closer AP.
Our main contributions are summarized as follows.
\begin{itemize}
    \item We formulate the energy minimization problem for WP-MEC networks with partial offloading.
        To the best of our knowledge, this is the first work that considers multiple WDs and multiple APs under online settings.
    \item We design an online algorithm by transforming the original problem into a series of deterministic optimization problems in each time slot
        based on the Lyapunov optimization theory.
        To reduce the time complexity of our algorithm, we propose a relax-then-adjust technique where
        the optimization problem is first relaxed and decomposed into several independent subproblems.
        After solving each subproblem, we adjust the computed values of variables to obtain a feasible solution.
    \item To solve the non-convex computation offloading subproblem obtained in the previous step, we propose an iterative algorithm based on the Alternating Minimization method.
        We demonstrate that each iterative step can be solved in polynomial time and the algorithm converges to local optima.
    \item Extensive simulations are conducted to validate the performance of our algorithm.
        Numerical results show that the total energy consumption is significantly reduced under various settings.
\end{itemize}

The rest of the paper is organized as follows.
In Section \ref{section:system model}, we introduce the system model and the problem formulation.
An online algorithm that jointly optimizes the WPT and computation scheduling is proposed in Section \ref{section:algorithm design}.
The simulations and related numerical results are presented in Section \ref{section:simulation} and we conclude our paper in Section \ref{section:conclusion}.

\section{System Model and Problem Formulation} \label{section:system model}
\begin{figure}[!t]
    \centering
    \includegraphics[width=0.45\textwidth]{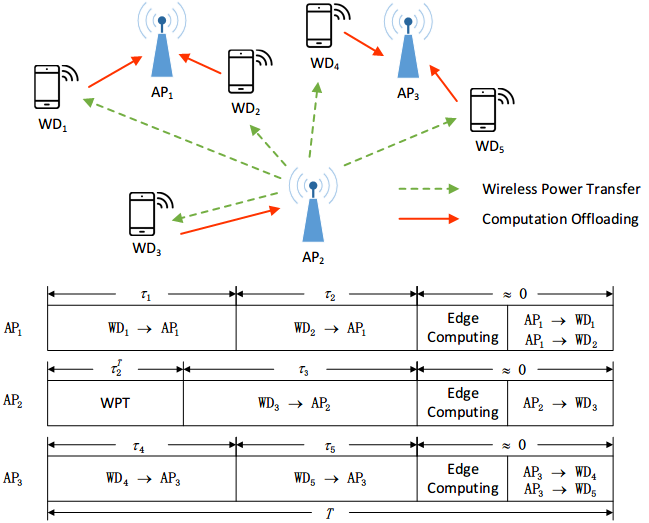}
    \caption{An example of system model and time allocation.}
    \label{fig:wpmec}
\end{figure}
As shown in Fig. \ref{fig:wpmec}, we consider a WP-MEC network consisting of
$N$ WDs and $M$ APs, where each AP is integrated with an RF energy transmitter and a MEC server.
APs are assumed to have a stable power supply and broadcast RF energy to WDs.
The energy harvested by each WD is stored in a rechargeable battery, which is used to power its computing and communication operations.
Similar to \cite{zhu2020computation}, we assume that WPT and wireless communications (for offloading) of the \emph{same} AP
cannot be performed simultaneously and the TDMA protocol is applied to avoid mutual interference,
but WPT and wireless communications of \emph{different} APs can be operated simultaneously over orthogonal frequency bands.

The time horizon is divided into slots with equal length $T$,
where each slot consists of four phases, i.e., WPT, computation offloading, edge computing, and result downloading,
as illustrated in Fig. \ref{fig:wpmec}.
Note that we have assumed WDs have simultaneous wireless information and power transfer (SWIPT) abilities, as shown in \cite{perera2017simultaneous}.
However, our model can be easily tailored for WDs without SWIPT capabilities, as discussed in Section \ref{subsection:problem_formulation}.
Since MEC servers have strong computation capacities compared with WDs and the computation results are of small data sizes, 
the time consumption for edge computing and result downloading is negligible \cite{zhu2020computation, wang2017joint, zhou2020computation, wu2019online, mao2019energy}.
Therefore, we only consider the duration of the first two phases in our model.
Let $h^D_{ij}(t)$ and $h^U_{ij}(t)$ denote the downlink and uplink channel gain between WD $i$ and AP $j$ on slot $t$.
If WD $i$ beyonds the communication scape of AP $j$, the corresponding channel gain is zero.
As in \cite{wang2020optimal, wu2019online, mao2019energy},
we assume all channels follow quasi-static flat-fading, i.e., the channel state remains constant within
each time slot, but may vary across different slots.
Our goal is to minimize the total energy consumption for processing the data of WDs.

\subsection{Wireless Power Transfer and Energy Harvesting Model}
According to \cite{naderi2014rf}, when multiple APs transfer RF power simultaneously, their energy waves may interfere with each other
and lead to a possible energy cancellation.
To improve the efficiency of RF energy transfer, we select only one AP to broadcast RF energy during each time slot.
By choosing different APs in turn, we also alleviate the ``double near-far'' effect because each WD has a chance to harvest energy from a close AP.
We use $a^T_j(t) \in \{0, 1\}$ to indicate whether AP $j$ broadcasts RF energy in slot $t$ and let $P^T_j(t)$ and $\tau^T_j(t)$ be the
corresponding transmission power and transmission time.
Then $a^T_j(t)$ should satisfy
\begin{equation}
    \sum_{j=1}^{M} a^T_j(t) \leq 1 \label{cons:a^T_j}
\end{equation}
and the energy consumption of AP $j$ during the WPT is
\begin{equation}
    E^T_j(t) = a^T_j(t) P^T_j(t) \tau^T_j(t). \label{eq:wpt_energy}
\end{equation}
As in \cite{mao2019energy, lu2014wireless}, we assume the energy harvested from noise is negligible and adopt a linear model to characterize the
energy harvesting circuit of WDs.
In particular, the energy harvested by WD $i$ during the $t$-th time slot is
\begin{equation}
    E^H_i(t) = \sum_{j=1}^{M} \mu_i a^T_j(t) P^T_j(t) h^D_{ij}(t) \tau^T_j(t) \label{eq:energy_harvest}
\end{equation}
where $\mu_j \in (0,1)$ is the energy conversion efficiency of WD $i$.
The harvested energy is stored in the battery of WDs.
Let $B^{max}_i$ and $B_i(t)$ be the capacity and remaining power of the battery in WD $i$,
then the update rule of $B_i(t)$ is
\begin{equation*}
    B_i(t+1) = \min \left[ B_i(t) - E^L_i(t) - E^O_i(t) + E^H_i(t), B^{max}_i \right]
\end{equation*}
where $E^L_i(t)$ and $E^O_i(t)$ are the energy consumption for local computation and wireless communication, as explained in the next subsection.
Due to the energy causality constraint, the energy consumption cannot exceed the available energy in the battery, so we must ensure
\begin{equation}
    E^L_i(t) + E^O_i(t) \leq B_i(t). \label{cons:energy_causality}
\end{equation}

\subsection{Computation Scheduling Model}
In each time slot $t$, let $A_i(t)$ be the amount of computation data arrived at WD $i$.
Without loss of generality, we assume $A_i(t)$ is i.i.d. with average rate $\mathbb{E}[A_i(t)] = \lambda_i$.
The arrived data can be either processed locally or offloaded to the APs.
The length of queueing data $Q_i(t)$ at WD $i$ evolves according to the following equation
\begin{equation*}
    Q_i(t+1) = Q_i(t) - D^L_i(t) - D^O_i(t) + A_i(t)
\end{equation*}
where $D^L_i(t)$ and $D^O_i(t)$ are the amount of locally processed data and offloaded data, respectively.
Note that the new data may arrive at the end of slot $t$ and cannot be processed until the beginning of next slot,
so $D^L_i(t)$ and $D^O_i(t)$ must satisfy
\begin{equation}
    D^L_i(t) + D^O_i(t) \leq Q_i(t). \label{cons:Q_i}
\end{equation}

\subsubsection{Local Computation}
As in previous researches, we assume the local computation and WPT can be performed simultaneously
and each WD $i$ adopts the Dynamic Voltage and Frequency Scaling technique \cite{rabaey2003digital} to control its CPU frequency $f_i(t)$.
Let $\tau^L_i(t)$ be the local computation time of WD $i$ in slot $t$, then we can express $D^L_i(t)$ with the following equation
\begin{equation}
    D^L_i(t) = \frac{f_i(t)\tau^L_i(t)}{\phi_i} \label{eq:def_D^L}
\end{equation}
where $\phi_i$ is the number of CPU cycles required to process one bit of computation data.
According to \cite{hu2018wireless}, the energy consumption for the local computation of WD $i$ is
\begin{equation}
    E^L_i(t) = \kappa_i f^3_i(t) \tau^L_i(t) \label{eq:def_E^L}
\end{equation}
where $\kappa_i$ is the energy efficiency coefficient of the chip equipped with WD $i$.
By substituting \eqref{eq:def_E^L} into \eqref{eq:def_D^L} we can obtain
\begin{equation*}
    D^L_i(t) = \sqrt[3]{\frac{E^L_i(t){\tau^L_i(t)}^2}{\kappa_i}} \cdot \frac{1}{\phi_i}
\end{equation*}
which is an increasing function with respect to the computation time $\tau^L_i(t)$ under fixed energy consumption $E^L_i(t)$.
Therefore, we can simply set
\begin{equation*}
    \tau^L_i(t) = T
\end{equation*}
to maximize the computation data processed locally.

\subsubsection{Computation Offloading}
Let $\tau_i(t)$ and $P_i(t)$ denote the offloading time and the transmit power of WD $i$, respectively.
Suppose each WD can only communicate with at most one AP within one time slot.
We use the binary variable $a_{ij}(t) \in \{0, 1\}$ to indicate whether WD $i$ is communicating with AP $j$,
then $a_{ij}(t)$ should satisfy
\begin{equation}
    \sum_{j=1}^{M} a_{ij}(t) \leq 1 \label{cons:a_ij}
\end{equation}
and the amount of computation data offloaded from WD $i$ to APs can be expressed as
\begin{equation*}
    D^O_i = \sum_{j=1}^{M} \frac{a_{ij}(t)B\tau_{i}(t)}{v_i} \log_2 \left( 1 + \frac{P_i(t) h^U_{ij}(t)}{\sigma_j^2} \right)
\end{equation*}
where $B$ is the spectrum bandwidth, $\sigma^2_j$ is the noise power of AP $j$, and
$v_i > 1$ indicates the communication overhead induced by encryption and packet header \cite{bi2018computation, zhou2020computation}.
The energy consumption for the computation offloading of WD $i$ is
\begin{equation*}
    E^O_i = \sum_{j=1}^{M} a_{ij}(t) P_i(t) \tau_i(t) = P_i(t) \tau_i(t).
\end{equation*}
Similar to \cite{wang2017joint}, we assume the energy consumption for the computation at AP $j$ is proportional to the total data it received from WDs
\begin{equation*}
    E^C_j(t) = \sum_{i=1}^{N} \eta a_{ij}(t) \phi_i D^O_i(t)
\end{equation*}
where $\eta$ is the energy consumption per CPU cycle of APs.

\subsection{Problem Formulation} \label{subsection:problem_formulation}
To reduce the carbon footprint,
in this paper, we aim to minimize the energy consumption for processing computation data arrived at WDs
by jointly optimizing the control decisions for WPT, local computation, and data offloading.
A similar objective is also considered in \cite{ji2018energy, mao2019energy, you2016energy}.
Based on our model, the considered problem can be formulated as follows.
\begin{alignat}{2}
    \min\quad & \lim_{H\to\infty} \frac{1}{H} \sum_{t=0}^{H-1} \sum_{j=1}^{M} \mathbb{E}\left\{ E^T_j(t) + E^C_j(t) \right\} &{}& \notag \\
    s.t.\quad & \eqref{cons:a^T_j}, \eqref{cons:energy_causality}, \eqref{cons:Q_i}, \eqref{cons:a_ij} &\quad& \notag \\
    & a^T_j(t)\tau^T_j(t) + \sum_{i=1}^{N} a_{ij}(t)\tau_i(t) \leq T &\quad& \forall j, \forall t \label{cons:time_allocation} \\
    & 0 \leq \tau^T_j(t) \leq T, 0 \leq P^T_j \leq P^{T, max}_j &\quad& \forall j, \forall t \label{cons:var_wpt} \\
    & 0 \leq \tau_{i}(t) \leq T, 0 \leq P_i(t) \leq P^{max}_i &\quad& \forall i, \forall t \label{cons:var_offloading} \\
    & 0 \leq f_i(t) \leq f^{max}_i &\quad& \forall i, \forall t \label{cons:var_local} \\
    & Q_i(t) \mbox{ is stable} &\quad& \forall i, \forall t \notag
\end{alignat}
where $P^{max}_i$ and $f^{max}_i$ is the maximum offloading power and computation capacity of WD $i$.
Constraint \eqref{cons:time_allocation} ensures the time allocation is feasible under the TDMA protocol.
The objective reflects APs' long-term energy consumption.
Since WDs harvest energy from APs, this is also the total energy consumption in the system.

For WDs without SWIPT capabilities, we only need to replace constraint \eqref{cons:time_allocation} with
\begin{equation*}
    \sum_{j'=1}^{M} a^T_{j'} \tau^T_{j'}(t) + \sum_{i=1}^{N} a_{ij}(t)\tau_i(t) \leq T \quad \forall j, \forall t
\end{equation*}
because the wireless communication is prohibited during WPT and thus the available communication time of \emph{all} APs (not just the one that transfers energy)
must exclude the WPT time.
It should be noted that the algorithm proposed in the next section can also be conveniently adapted to this situation.

\section{An Online Algorithm for Wireless Power Transfer and Computation Scheduling} \label{section:algorithm design}

In this section, we design an online algorithm that jointly optimizes WPT and computation scheduling based on the Lyapunov optimization.
A relax-then-adjust technique is proposed to decompose the original problem into smaller subproblems so that the algorithm's complexity
is significantly reduced.

\subsection{Algorithm Design with Lyapunov Optimization}
For convenience of description, we first define the battery shortage $B^-_i(t)$ as
\begin{equation*}
    B^-_i(t) = B^{max}_i - B_i(t).
\end{equation*}
According to the constraint \eqref{cons:energy_causality}, the energy consumed by WD $i$ on slot $t$ cannot exceed the remaining energy in its battery.
From a long-term perspective, this implies the time-average harvested energy by WD $i$ is equal to or greater than its time-average energy consumption,
so $B^-_i(t)$ is also stable in the long run.
Let $\bm{\Theta}(t) = [ \bm{Q}(t), \bm{B}^-(t) ]$ be the combined queue vector, where $\bm{Q}(t) = (Q_1(t), Q_2(t), \dots, Q_N(t))$ and
$\bm{B}^-(t) = (B^-_1(t), B^-_2(t), \dots, B^-_N(t))$.
According to the Lyapunov optimization theory, we start by defining the quadratic Lyapunov function
\begin{equation*}
    L(t) = \frac{1}{2} \sum_{i=1}^{N} \left[ Q_i(t)^2 + B^-_i(t)^2 \right]
\end{equation*}
and the conditional Lyapunov drift
\begin{equation*}
    \Delta L(t) = \mathbb{E} \left\{ L(t+1) - L(t) | \bm{\Theta}(t) \right\}.
\end{equation*}
Next, we combine $\Delta L(t)$ with the objective function and form the following drift-plus-penalty term
\begin{equation*}
    \Delta_V L(t) = \Delta L(t) + V\mathbb{E} \left\{ \sum_{j=1}^{M} \left( E^T_j(t) + E^C_j(t) \right) | \bm{\Theta}(t) \right\}
\end{equation*}
where $V$ is a tunable parameter that controls the trade-off between the energy consumption and the queueing delay of computation data.
The following lemma provides an upper bound for $\Delta_V L(t)$.
\begin{lemma}
    On every slot $t$ and for any value of $\bm{\Theta}(t)$, the drift-plus-penalty term always satisfies
    \begin{align}
        \Delta_V L(t) \leq &C - \sum_{i=1}^{N} Q_i(t) \mathbb{E} \left\{ D^L_i(t) + D^O_i(t) - A_i(t) | \bm{\Theta}(t) \right\} \notag \\
        &- \sum_{i=1}^{N} B^-_i(t) \mathbb{E} \left\{ E^H_i(t) - E^L_i(t) - E^O_i(t) | \bm{\Theta}(t) \right\} \notag \\
        &+ V\mathbb{E} \left\{ \sum_{j=1}^{M} \left( E^T_j(t) + E^C_j(t) \right) | \bm{\Theta}(t) \right\} \label{eq:dpp_bound}
    \end{align}
    where $C$ is a constant defined in the proof.
    \label{lemma:dpp_bound}
\end{lemma}

The proof follows a standard procedure \cite{neely2010stochastic} and is omitted for brevity.
Based on the Lyapunov optimization theory, we can obtain an approximately optimal algorithm of our problem
by minimizing the right-hand side of \eqref{eq:dpp_bound} in every slot $t$.
The resulting time-average energy consumption decreases at the rate of $O(1/V)$ and
the time-average queueing delay increases at the rate of $O(V)$, presenting a $O(1/V)$-$O(V)$ tradeoff between the two metrics.
As a result, we can approach the optimal energy consumption arbitrarily close by increasing the value of $V$.
However, in our problem, directly solving the minimization problem is difficult because the decision variables are coupled in constraints such as \eqref{cons:energy_causality} and \eqref{cons:Q_i}.
To reduce the complexity of our algorithm, we propose a relax-then-adjust technique to decompose the original minimization problem into independent subproblems.
The details are presented in the next subsection.

\subsection{Relax-Then-Adjust}
In our problem, the objective is to minimize the system energy consumption used to process the computation data of WDs.
When the amount of arrived workload is fixed, this is equivalent to maximizing the energy efficiency, which is defined as
the ratio of total energy consumption to the corresponding aggregate accomplished computation data.
Inspired by this interpretation, we define the \emph{marginal energy efficiency} of local computation and computation offloading,
denoted by $\epsilon^L_i(t)$ and $\epsilon^O_i(t)$ respectively, as follows
\begin{align}
    \epsilon^L_i(t) &= \frac{\partial E^L_i(t)}{\partial D^L_i(t)} = 3\kappa_i \phi_i f^2_i(t) \notag \\
    \epsilon^O_i(t) &= \frac{\partial \left( E^O_i(t) + \eta \phi_i D^O_i(t) \right)}{\partial D^O_i(t)} \notag \\
    &= \sum_{j=1}^{M} \frac{a_{ij}(t)v_i \ln 2}{B} \cdot \left( \frac{\sigma^2_j}{h^U_{ij}(t)} + P_i(t) \right) + \eta\phi_i \label{def:offloading_ee}
\end{align}
where we have assumed the time allocation $\tau_{i}(t)$ is given in deriving \eqref{def:offloading_ee}.
According to the optimality conditions, the two marginal energy efficiency should be equal when the overall energy efficiency is maximized.
Based on this fact, we can first relax constraint \eqref{cons:energy_causality} and \eqref{cons:Q_i} and then
adjust the values of $f_i(t)$ and $P_i(t)$ to obtain a feasible solution that makes $\epsilon^L_i(t) = \epsilon^O_i(t)$.
A similar technique is also used to decouple the WPT and computation offloading variables in constraint \eqref{cons:time_allocation}.
After relaxing constraints \eqref{cons:energy_causality}, \eqref{cons:Q_i}, and \eqref{cons:time_allocation}, the optimization problem of minimizing the right-hand side
of \eqref{eq:dpp_bound} can be decomposed into three subproblems by classifying variables into independent groups.
In our algorithm, we first compute solutions for these subproblems and then adjust their values according to the optimality conditions.
The details of our algorithm are described as follows.

\subsubsection{WPT}
We compute the WPT-related control variables by solving the following subproblem
\begin{alignat}{2}
    \min\quad & \sum_{j=1}^{M} \left( V - \sum_{i=1}^{N} B^-_i(t)\mu_i h^D_{ij}(t) \right) a^T_j(t) P^T_j(t) \tau^T_j(t)  &{}& \label{obj:wpt}
\end{alignat}
subject to constraints \eqref{cons:a^T_j} and \eqref{cons:var_wpt}.
Note that we have relaxed the time allocation constraint \eqref{cons:time_allocation}.
This problem is derived by substituting \eqref{eq:wpt_energy} and \eqref{eq:energy_harvest} into \eqref{eq:dpp_bound}
and group WPT-related terms together.
The next two subproblems are derived in a similar way.
Let $c^T_j(t) = V - \sum_{i=1}^{N} B^-_i(t)\mu_i h^D_{ij}(t)$ be the coefficient of $a^T_j(t) P^T_j(t) \tau^T_j(t)$.
If $c^T_j(t) \geq 0$ for all $j$, then the optimal value is obtained by setting all $a^T_j(t)$ to $0$.
Otherwise, find $j^* = \argmin c^T_j(t) P^{T, max}_j$ and the solution is $a^T_{j^*}(t) = 1, P^T_{j^*}(t) = P^{T,max}_{j^*}, \tau^T_{j^*}(t) = T$.

\subsubsection{Local Computation} \label{section:local_computation}
The only variable for local computation is the CPU frequency of WDs.
By relaxing constraints \eqref{cons:energy_causality} and \eqref{cons:Q_i}, we get the following subproblem
\begin{equation}
    \min \sum_{i=1}^{N} B^-_i(t) \kappa_i f^3_i(t) T - \sum_{i=1}^{N} Q_i(t) \frac{f_i(t) T}{\phi_i} \label{subproblem:local}
\end{equation}
where $f_i(t)$ is subject to constraint \eqref{cons:var_local}.
This problem can be further divided into $N$ subproblems because the variable $f_i(t)$ is independent with each other.
As a result, the problem can be solved analytically and
the optimal value of $f_i(t)$ is obtained at either the boundary points or the stationary point of \eqref{subproblem:local},
which is given by
\begin{equation*}
    f_i(t) = \min \left\{ f^{max}_i, \sqrt{\frac{Q_i(t)}{3\kappa_i \phi_i B^-_i(t)}} \right\}.
\end{equation*}

\subsubsection{Computation Offloading} \label{step:computation_offloading}
The optimal time allocation of computation offloading can be derived by solving
\begin{alignat}{2}
    \min\quad & \sum_{j=1}^{M} \sum_{i=1}^{N} V\eta a_{ij}(t)\phi_i D^O_i(t) - \sum_{i=1}^{N} Q_i(t)D^O_i(t) &{}& \notag \\
    & + \sum_{i=1}^{N} B^-_i(t) P_i(t) \tau_i(t) &{}& \label{obj:offloading} \\
    s.t.\quad & \eqref{cons:a_ij}, \eqref{cons:var_offloading} &{}& \notag \\
    & \sum_{i=1}^{N} a_{ij}(t)\tau_i(t) \leq T  \qquad \forall j, \forall t &{}& \label{cons:relax_time_allocation} 
\end{alignat}
where we have used a relaxed time allocation constraint \eqref{cons:relax_time_allocation} instead of \eqref{cons:time_allocation}.
This MIP problem is non-convex and has no efficient algorithms in general.
In the next subsection, we will devise a heuristic algorithm for this problem based on the Alternating Minimization method.

\subsubsection{Adjust Variable Values}
If the variable values computed in previous steps satisfy constraint \eqref{cons:energy_causality}, \eqref{cons:Q_i}, and \eqref{cons:time_allocation},
then we can skip this step.
Otherwise, we have to adjust their values to obtain a feasible solution.
As described earlier, our intuition is to equalize the marginal benefits of variables constrained by the same resource.

Let us first consider the case where WDs have sufficient queueing data so constraint \eqref{cons:Q_i} is redundant.
Without loss of generality, we can assume the energy causality constraint \eqref{cons:energy_causality} is tight.
Let $\epsilon^L_i(t) = \epsilon^O_i(t)$,
we can express the local CPU frequency $f_i(t)$ with respect to the offloading power $P_i(t)$
\begin{equation}
    f_i(t) = \sqrt{\sum_{j=1}^{M} \frac{a_{ij}(t)v_i\ln 2}{3\kappa_i \phi_i B} \cdot \left( \frac{\sigma^2_j}{h^U_{ij}(t)} + P_i(t) \right) + \frac{\eta}{3\kappa_i}}. \label{eq:f_i_adjust}
\end{equation}
Suppose AP $j^*$ is the one that broadcasts RF energy in time slot $t$
and let $\mathcal{N}^*(t)$ denote the set of WDs that offload data to AP $j^*$, i.e., $a_{ij^*}(t) = 1$ and $\tau_i(t) > 0$ for all $i\in \mathcal{N}^*(t)$.
If WD $i$ does not belong to $\mathcal{N}^*(t)$, which means it communicates with some other AP $j$ such that $a^T_j = 0$,
then the time allocation constraint \eqref{cons:time_allocation} of $\tau_i(t)$ reduces to \eqref{cons:relax_time_allocation}.
Therefore, the time allocation computed in Section \ref{step:computation_offloading} is feasible and no need for adjustment.
Thus, we only need to adjust the values of $f_i(t)$ and $P_i(t)$ by solving
\begin{equation}
    E^L_i(t) + E^O_i(t) = B_i(t), \label{eq:energy_causality}
\end{equation}
which results in
\begin{equation}
    P_i(t) = \frac{B_i(t) - \kappa_i f^3_i(t) T}{\tau_i(t)}. \label{eq:P_vs_tau}
\end{equation}
For the rest WDs belongs to $\mathcal{N}^*(t)$, we first re-allocate the WPT time $\tau^T_{j^*}(t)$ and offloading time $\tau_i(t), i\in \mathcal{N}^*(t)$.
According to \eqref{obj:wpt} and \eqref{obj:offloading}, the marginal cost of $\tau^T_{j^*}(t)$ is $c^T_{j^*}(t)P^{T, max}_{j^*}$
and the marginal cost of $\tau_i(t)$ for all $i\in \mathcal{N}^*(t)$ is
\begin{align}
    &\frac{\partial \left( V\eta\phi_i D^O_i(t) - Q_i(t)D^O_i(t) + B^-_i(t) P_i(t) \tau_i(t) \right)}{\partial \tau_i(t)} \notag \\
    = & \frac{\left( V\eta\phi_i - Q_i(t) \right)B}{v_i} \log_2 \left( 1 + \frac{P_i(t)h^U_{ij^*}(t)}{\sigma^2_{j^*}} \right) + B^-_i(t)P_i(t). \label{eq:marginal_cost_tau}
\end{align}
By substituting \eqref{eq:P_vs_tau} into \eqref{eq:marginal_cost_tau}, the marginal cost of $\tau_i(t)$ is a function of itself.
Due to the optimality condition, we can compute the new time allocation by equalizing the marginal cost of $\tau_i(t)$ and $\tau^T_{j^*}(t)$.
After that, the values of $P_i(t)$ and $f_i(t)$ for WDs in $\mathcal{N}^*(t)$ are adjusted just like other WDs.

For cases where the remaining energy in the battery is adequate to handle all the queueing data in WDs,
the data constraint \eqref{cons:Q_i} is tight and the energy constraint \eqref{cons:energy_causality} is redundant.
As a result, we can repeat the above procedures by replacing \eqref{eq:energy_causality} with
\begin{equation*}
    D^L_i(t) + D^O_i(t) = Q_i(t).
\end{equation*}


\subsection{A Heuristic Algorithm for Computation Offloading}
In this subsection, we propose a heuristic algorithm for the optimization problem of computation offloading based on the Alternating Minimization (AM) method.
The main idea underlying AM is to replace the difficult joint optimization with a sequence of easier optimization involving grouped subsets of the variables.
In our problem, we partition the decision variables into two groups: the time allocation variables $a_{ij}(t)\tau_i(t)$ and the transmission power variables $P_i(t)$.
According to the AM method, the original problem can be solved iteratively by solving the following subproblems in each step $k$
\begin{align}
    P^k(t) = \argmin_{P(t)} g\left( a^k(t)\tau^k(t), P(t) \right) \label{subproblem:offloading_power} \\
    a^{k+1}(t)\tau^{k+1}(t) = \argmin_{a(t)\tau(t)} g\left( a(t)\tau(t), P^k(t) \right) \label{subproblem:offloading_time_allocation}
\end{align}
where $g(t)$ is the objective function in \eqref{obj:offloading}.
Theoretical analysis guarantees that the computed solution converges to local minima \cite{bezdek2003convergence}.

\subsubsection{Transmission Power Selection}
When the time allocation is given, minimizing \eqref{obj:offloading} can be decomposed into $N$ independent subproblems
and we can obtain an analytical solution as in Section \ref{section:local_computation}.
If $\phi_i \eta > Q_i(t)$, then the optimal value of $P_i(t)$ is $P^{max}_i$.
Otherwise, set 
\begin{equation*}
    P_i(t) = \min \left\{P^{max}_i, \frac{( Q_i(t) - \phi_i\eta )B}{B^-_i(t) v_i \ln 2} - \frac{\sigma^2_{j'}}{h^U_{ij'}(t)} \right\}
\end{equation*}
where $j'$ is the AP that WD $i$ communicate with.





\subsubsection{Time Allocation}
Since the objective function is linear with respect to $\tau_i(t)$, the marginal cost of time allocation is constant.
According to the optimality condition, there is an optimal solution in which each AP $j$ is devoted to at most one WD $j$.
Therefore, we can assume $\tau_i(t) = T$ without loss of generality.
Then the problem is to determine the value of $a_{ij}(t)$, which turns to be a standard assignment problem and
can be solved within $O(M^2N + N^2\log N)$ by the Hungarian algorithm \cite{kuhn1955hungarian}.

\section{Simulation Results} \label{section:simulation}
In this section, we evaluate the proposed algorithm through simulations and compare its performance with the following two benchmark schemes:
\begin{itemize}
    \item \emph{Local Computation Only (LCO) scheme:} the computation data of WDs are processed locally;
    \item \emph{Fully Offloading (FO) scheme:} the computation data of WDs are fully offloaded to the APs.
\end{itemize}
Similar to our algorithm, we optimize the control decisions for these two schemes based on the Lyapunov optimization, 
so their performance is also associated with the parameter $V$.

We consider a WPMEC network with $N=30$ WDs and $M=5$ APs.
The simulation settings are selected based on the works in \cite{mao2019energy} and \cite{zhu2020computation}.
The maximum CPU frequency and battery capacity of each WD are $f^{max}_i = 0.5GHz$ and $B^{max}_i = 30kJ$, respectively.
As in \cite{zhu2020computation}, we adopt a simplified Rayleigh fading channel model and the uplink gain from WD $i$ to AP $j$
is $h^U_{ij} = \theta^U d^{-3}_{ij} \bar{h}_{ij}$, where $\theta^U = 6.25 \times 10^{-4}$ (i.e. $-32dB$), $d_{ij}$ is the distance between WD $i$
and AP $j$, and $\bar{h}_{ij}$ is a random variable drawn from the standard complex normal distribution $\mathcal{CN}(0,1)$.
We set the downlink gain as $h^D_{ij} = 2h^U_{ij}$.
The other parameters used in our simulations are $B = 1MHz$, $\mu_i = 0.51$, $\kappa_i = 10^{-28}$, $v_i = 1.1$, $\eta = 8.2nJ$, $\phi_i = 1000\ cycles/bit$,
and $\sigma^2_j = 10^{-9}W$.

\begin{figure}[t]
\begin{minipage}[t]{0.49\linewidth}
\centering
\includegraphics[width=1.7in]{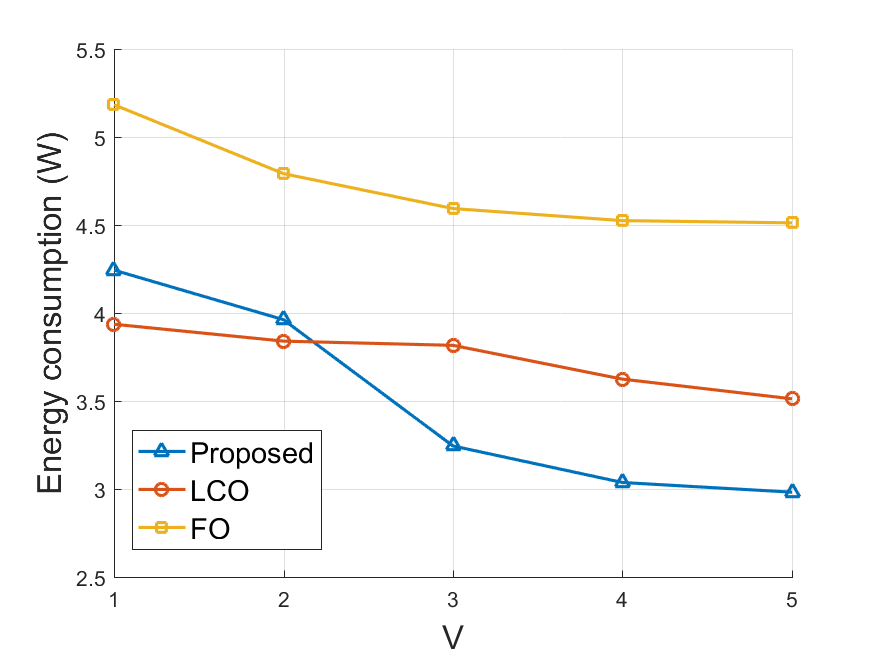}
\caption{System energy consumption vs. parameter $V$.}
\label{fig:energy}
\end{minipage}
\begin{minipage}[t]{0.49\linewidth}
\centering
\includegraphics[width=1.7in]{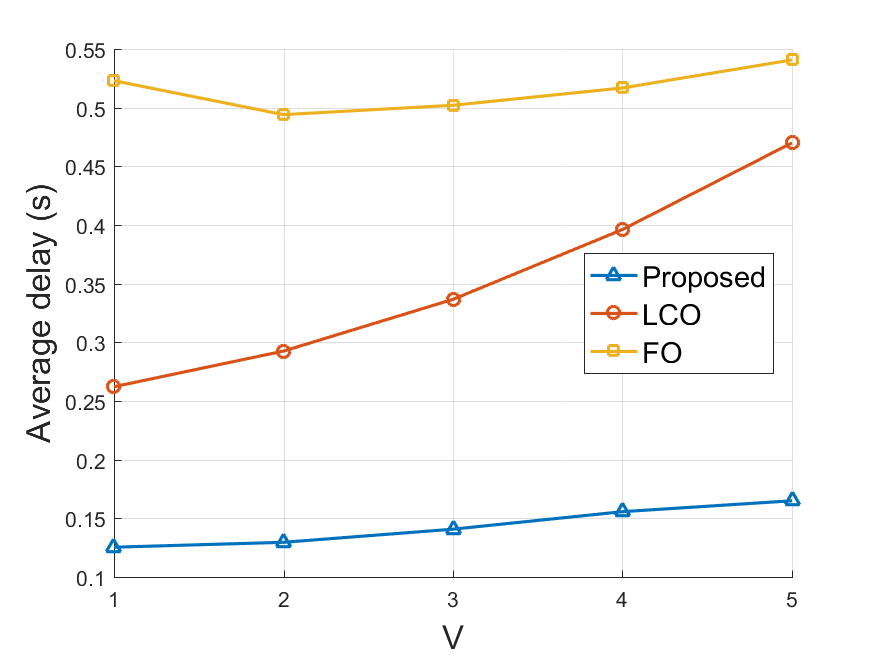}
\caption{Average delay vs. parameter $V$.}
\label{fig:delay}
\end{minipage}
\end{figure}

The impact of control parameter $V$ on the system energy consumption and average delay are demonstrated in Fig. \ref{fig:energy}
and Fig. \ref{fig:delay}, respectively.
In all algorithms, the energy consumption decreases with $V$ and the average delay grows with $V$, which is in accordance with the energy-delay
tradeoff of Lyapunov optimization.
As shown in Fig. \ref{fig:energy}, the energy consumption of the proposed algorithm is slightly higher than LCO when $V$ is small,
but is significantly better than the other two when $V$ is large.
Meanwhile, the average delay induced by our algorithm outperforms the two benchmarks in all cases.
This is because our algorithm performs local computation and computation offloading simultaneously, thus resulting in
higher process rate and better energy efficiency.

\begin{figure}[h]
\begin{minipage}[t]{0.49\linewidth}
\centering
\includegraphics[width=1.7in]{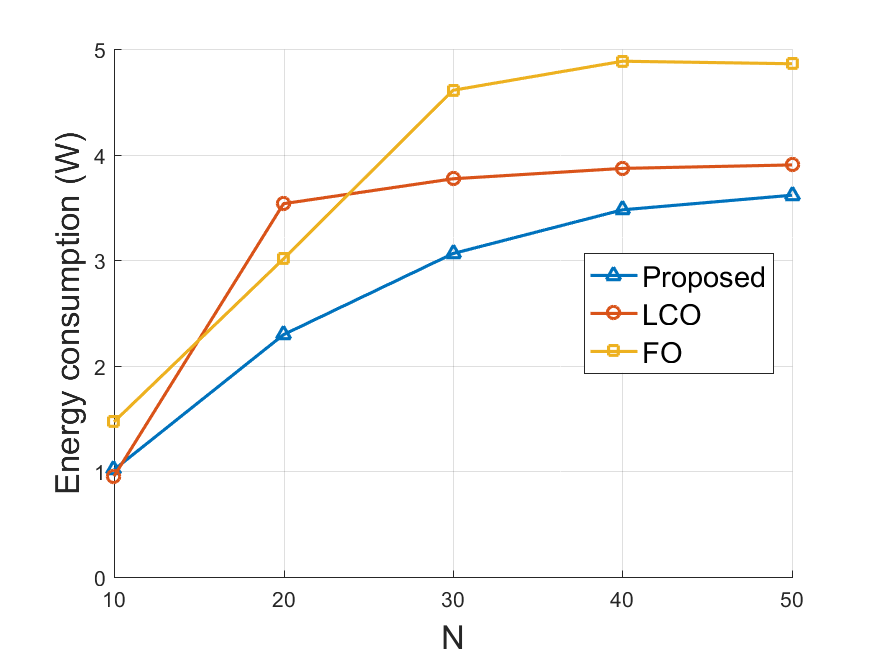}
\caption{System energy consumption vs. number of WDs $N$.}
\label{fig:energy_change_n}
\end{minipage}
\begin{minipage}[t]{0.49\linewidth}
\centering
\includegraphics[width=1.7in]{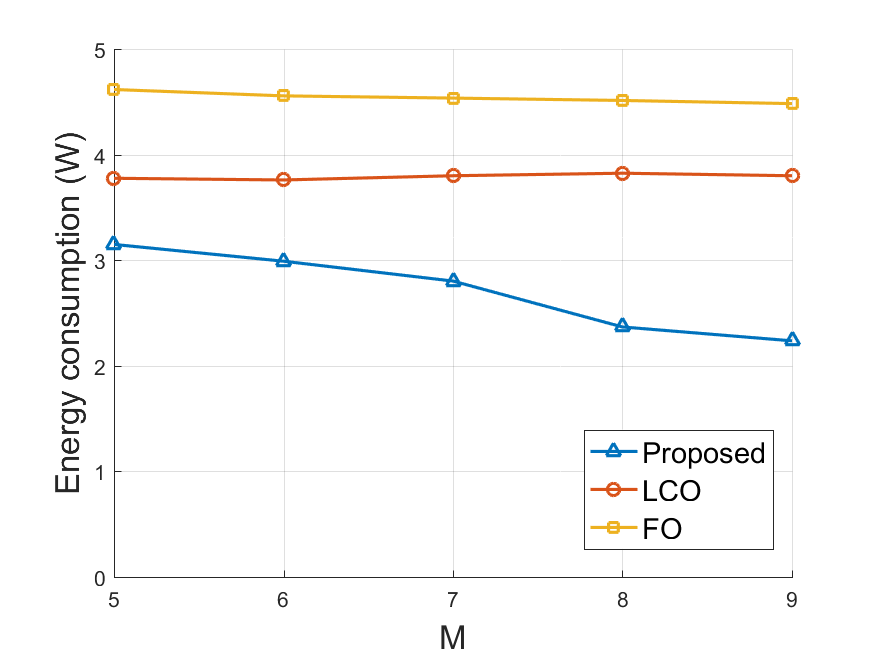}
\caption{System energy consumption vs. number of APs $M$.}
\label{fig:energy_change_m}
\end{minipage}
\vspace{5pt}
\end{figure}

In addition to the varying parameter $V$, we also conducted simulations under distinct network scales.
The system energy consumption under different number of WDs and different number of APs are presented in Fig. \ref{fig:energy_change_n}
and Fig. \ref{fig:energy_change_m}.
The default value of $V$ is set to $3$ in all simulations.
The growth of $N$ means there are more data to be processed in the system, thus results in higher energy consumption.
When $N=40$, APs already operate at the peak WPT power in FO but we still observed an extremely large average delay.
We analysized the run-time data and found that the generated workload exceeds the system capacity.
This also explains why the energy consumption of FO does not increase when $N$ grows from $40$ to $50$.
In Fig. \ref{fig:energy_change_m}, the energy consumption of FO decreases with respect to $M$ because WDs can choose closer APs
to offload their computation tasks.
However, the improvement on LCO is very limited.
In both situations, our algorithm can utilize the advantages of both local computation and computation offloading,
thus achieves the minimal energy consumption and average delay.

\section{Conclusions} \label{section:conclusion}
In this paper, we investigate the computation scheduling problem in WP-MEC networks with multiple WDs and multiple APs.
To minimize energy consumption, we propose an online algorithm that jointly optimizes the resource allocation in WPT and computation scheduling.
The time complexity of our algorithm is significantly reduced via the relax-then-adjust technique and the Alternating Minimization method.
Simulations are conducted to validate the performance of our algorithm.

\bibliographystyle{IEEEtran}
\bibliography{ref}

\end{document}